\documentclass[nofootinbib,aps,amsfonts,superscriptaddress]{revtex4}
\usepackage{amsmath}
\usepackage{epsfig,epsf}
\usepackage{graphicx}
\usepackage{verbatim}
\usepackage{epstopdf}
\usepackage{epstopdf}

\def\beq{\begin{equation}}
\def\eeq{\end{equation}}
\def\bea{\begin{eqnarray}}
\def\eea{\end{eqnarray}}
\def\eq#1{{Eq.~(\ref{#1})}}
\def\fig#1{{Fig.~\ref{#1}}}

\newcommand{\Lb}{\left(}
\newcommand{\Rb}{\right)}

\def\pom{{I\!\!P}}
\def\reg{{I\!\!R}}
\begin{document}


\voffset1.5cm
\title{Inclusive production in a QCD and N=4 SYM motivated model for soft
interactions}
\author{Errol Gotsman}
\affiliation{Department of Particle Physics, Tel Aviv University, 
Tel Aviv 69978, Israel}
\author{Eugene Levin}
\affiliation{Department of Particle Physics, Tel Aviv University, 
Tel Aviv 69978, Israel}
\affiliation{Departamento de F\'\i sica, 
Universidad T\'ecnica Federico Santa Mar\'\i a, Avda. Espa\~na 1680,
Casilla 110-V,  Valparaiso, Chile} 
\author{Uri Maor}
\affiliation{Department of Particle Physics, Tel Aviv University, 
Tel Aviv 69978, Israel}

\date{\today}
\begin{abstract}
The results presented in this paper differ from our previous
unsuccessful attempt to predict the rapidity distribution
at $W = 7 \,TeV$. The original version of our model
(GLMM) only summed a particular class of Pomeron diagrams
(enhanced diagrams). We believe that this  was the reason  
for our failure to describe the $7 \,TeV$ inclusive LHC data.
We  have developed a new approach (GLM) that also
includes the summation of the semi-enhanced diagrams.
This contribution is essential for a successful
description of the inclusive distributions, which is presented here.
\end{abstract}
\maketitle

Traditionally, inclusive hadron production at high energies has
been considered a typical example of a soft process.
Due to our lack of understanding of QCD at long distances,
such processes are studied in
the framework of high energy phenomenology based on soft
Pomerons and their interactions. 
However, the
first LHC data on hadron production \cite{ALICE, CMS,ATLAS} showed
that the alternative approach, based on high density 
QCD\cite{GLR,MUQI,MV,B,K,JIMWLK,KLN,KLNLHC},
is able to predict\cite{LRPP} and describe the main features
of the inclusive experimental data at the LHC \cite{LR,MCLPR,PRA}.
On the other hand, our first attempt
\cite{GLMINC} to predict the inclusive rapidity spectra based on 
a model of soft interactions\cite{GLMM} 
failed to describe the experimental
data, possibly giving the impression that 
soft models are unable to depict the LHC data.
In this letter we show that this
impression is incorrect, and that
models based on soft Pomeron interactions are 
capable of reproducing the inclusive LHC data. In our two papers
(see Refs. \cite{GLMM,GLMLAST}) we have built 
a model for soft interactions that is based 
on two theoretical precepts:
high energy scattering in N=4 SYM, which stems from Pomeron interactions;
and its matching with 
perturbative QCD, where we can use the BFKL Pomeron
calculus to obtain the scattering amplitude. 
Our model includes: 

(i) Pomeron $\Delta_\pom \approx $  0.2 and $\alpha'_\pom \approx $ 0.
The large intercept 
of the Pomeron follows from N=4 SYM, and its value 
is compatible with N=4 SYM models for DIS data\cite{LEPO}.

(ii) A large Good-Walker component as in N=4 SYM, 
where the main contributions are due
to elastic scattering and diffraction production \cite{BST,HIM}. 

(iii) Weak Pomeron interaction, which is of the order of
$2/\sqrt{\lambda} \ll 1$ in N=4 SYM.

(iv) Only the triple Pomeron vertex is necessary to provide
a natural matching with perturbative QCD.

In our recent paper\cite{GLMLAST} we summed 
all essential Pomeron diagrams. 
Therefore, it is important for consistency,   
to check whether our postulates are necessary for 
the description of 
the inclusive hadron production at the LHC.

\begin{figure}
\centerline{\epsfig{file=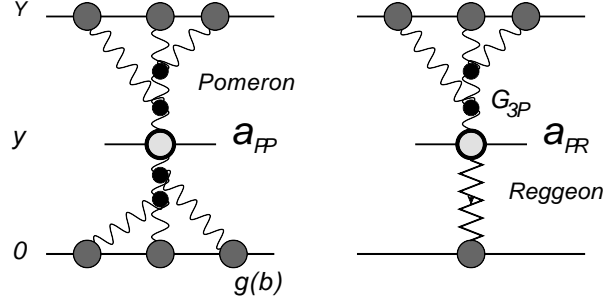,width=80mm}}
\caption{The Mueller diagrams \cite{MUDI} for single inclusive cross section.
The wavy bold line denotes the exact Pomeron Green's function of \eq{IXS2}, 
which is the sum of the enhanced diagrams. 
The zig-zag line stands for the exchange of a Reggeon.}
\label{incldi}
\end{figure}

The inclusive cross section can be calculated in 
Pomeron calculus\cite{GRIBRT}
(see also Refs.\cite{COL,SOFT,LEREG,AGK})
using Mueller diagrams\cite{MUDI} shown in \fig{incldi}.  
They lead to the following expression for the single inclusive cross section
\bea \label{IXS7}
&&\frac{1}{\sigma_{NSD}}\,\frac{d \sigma}{d y}\,\,
=\,\,\frac{1}{\sigma_{NSD}(Y)}\,\left\{ a_{\pom \pom}
\Big( \int d^2 b \Big( \alpha^2 \,G_1(b, Y/2 - y) 
+ \beta^2 G_2(b,Y/2 -y)\Big)\right. \nonumber\\
&&\left.\times\,\int d^2 b \Big( \alpha^2 \,G_1(b, Y/2 + y) 
+ \beta^2 G_2(b,Y/2 +y)\Big)  \right.\\
&&\left.\,\,\,-\,\,a_{ \pom \reg} \,\,( \alpha^2 \,g^\reg_1 
+ \beta^2 g^\reg_2)\,\Big[\alpha^2 \,
\int d^2 b \Big( \alpha^2 \,G_1(b, Y/2 - y) 
+ \beta^2 G_2(b,Y/2 -y)\Big)\,e^{\Delta_\reg\,(Y/2 + y)}\right.\nonumber\\
&&\left.+\,\int d^2 b \Big( \alpha^2 \,G_1(b, Y/2 + y) 
+ \beta^2 G_2(b,Y/2 +y)\Big)\,e^{\Delta_\reg\,(Y/2 - y)} 
\Big]\right\}.\nonumber
\eea
In \eq{IXS7} $G_i\Lb b,Y\Rb$ denotes the sum of 'fan' diagrams
\beq \label{IXS8}
G_i\Lb b ,Y\Rb\,\,=\,\,\Lb g_i\Lb b \Rb/\gamma\Rb\,
G_{enh}(y)/\Big(1 + \Lb G_{3\pom}/\gamma\Rb\,g_i\Lb b 
\Rb\,G_{enh}(y)\Big),
\eeq
where the Green's function of the Pomeron, obtained by the summation 
of the enhanced diagrams\cite{GLMM}, is equal to
\beq \label{IXS2}
G_{enh}\Lb Y\Rb\,\,=\,\,1 \,
-\,\exp\Lb \frac{1}{T\Lb Y\Rb}\Rb\,
\frac{1}{T\Lb Y\Rb}\,\Gamma\Lb 0,\frac{1}{T\Lb Y\Rb} \Rb.
\eeq

In \eq{IXS8} and \eq{IXS2} we denote 
(see Refs.\cite{GLMM,GLMLAST} for details):
\beq \label{NO}
g_i\Lb b \Rb\,=\,g_i S_i\Lb b \Rb\,
=\,\,\frac{g_i}{4 \pi}\,m^3_i\,b\,K_1\Lb m_i\,b\Rb;\,\,T\Lb Y\Rb\,\,\,
=\,\,\gamma e^{\Delta_{\pom}\,Y};
\,\,\,\,\gamma^2\,=\,\,\int d^2 k G_{3\pom}. 
\eeq
$g_i$ denotes the Pomeron vertex of interaction with the 
$i$ state at b=0; $m_i$ is the parameter
which determines the impact parameter dependence of this vertex;
$\Delta_\pom$ and $\Delta_{\reg}$ are the Pomeron and Reggeon 
trajectory intercepts respectively. 
The triple Pomeron vertex $G_{3\pom}$ and
the parameters $\alpha$ and $\beta$, which determine the 
decomposition of the proton wave function into its GW components,
$\Psi_{proton} = \alpha \Psi_1 + \beta \Psi_2$, have been discussed 
in Refs.\cite{GLMM,GLMLAST}. 
In our calculations we took the 
numerical values of these parameters from Ref.\cite{GLMLAST}.

In \eq{IXS7} we introduce two new vertices:
$a_{\pom \pom}$ and $a_{\pom \reg}$,
which describe the emission of hadrons from
Pomeron and from the secondary Reggeon (see \fig{incldi}).
In practice, we have to deal with two more
dimensional parameters $Q$ and $Q_0$.
$Q$ is the average transverse momentum of the produced minijets,
and $Q_0/2$ is the mass
of the slowest hadron produced in the decay of the minijet.
As we shall see below, the
only parameter that determines the inclusive spectra is the ratio $Q_0/Q$.

We need these parameters to calculate the pseudo-rapidity $\eta$,
which we use instead of the rapidity $y$. 
The relation between $y$ ans $\eta$ is well known (see Ref.\cite{KL}) 
\beq \label{IXS3}
y\Lb \eta, Q_0/Q\Rb \,\,=
\,\,\frac{1}{2} \ln\left\{\frac{\sqrt{\frac{Q_0}{Q} 
+ 1\,+\,\sinh^2\eta}\,\,+\,\,\sinh \eta}{\sqrt{\frac{Q_0 }{Q} 
+1 \,+\,\sinh^2\eta}\,\,-\,\,\sinh \eta}\right\},
\eeq
with the Jacobian
\beq \label{IXS4}
h\Lb \eta,Q_0/Q\Rb \,\,= \frac{\cosh\eta}{\sqrt{\frac{Q_0}{Q}\,
+\,1\,+\,\sinh^2\eta}}. 
\eeq

Using \eq{IXS3} and \eq{IXS4} we can re-write \eq{IXS7} in the form
\bea \label{IXS9}
&&\frac{1}{\sigma_{NSD}}\,\frac{d \sigma}{d \eta}\,\,
=\,\,h\Lb \eta,Q_0/Q\Rb \frac{1}{\sigma_{NSD}(Y)}\,
\left\{ a_{\pom \pom}\left( \int d^2 b \Big\{\alpha^2 \,G_1
\Big(b, Y/2 - y\Lb \eta, Q_0/Q\Rb\Big)    
+ \beta^2 G_2\Big(b, Y/2 - y\Lb 
\eta, Q_0/Q\Rb\Big)\Big\}\right.\right. \nonumber\\
&&\left.\left.\times\,\int d^2 b 
\Big\{ \alpha^2 \,G_1\Big(b, Y/2 + y\Lb \eta, Q_0/Q\Rb\Big) 
+ \beta^2 \,G_1\Big(b, Y/2 + y\Lb \eta, Q_0/Q\Rb\Big)\Big\}\right)  
\right.\\
&&\left.\,\,\,-\,\,a_{\pom \reg} \,\,( \alpha^2 \,g^R_1 
+ \beta^2 g^R_2)\,\Big[\alpha^2 \,\int d^2 b 
\Big\{ \alpha^2 \,G_1\Big(b, Y/2 - y\Lb \eta, Q_0/Q\Rb\Big)  
+ \beta^2 G_2\Big(b, Y/2 - y\Lb \eta, Q_0/Q\Rb\Big)\Big\}\,
e^{\Delta_R\,(Y/2 + y)}\right.\nonumber\\
&&\left.+\,\int d^2 b \Big\{\alpha^2 \,G_1
\Big(b, Y/2 + y\Lb \eta, Q_0/Q\Rb\Big) 
+ \beta^2 G_2\Big(b, Y/2 +y\Lb \eta, Q_0/Q\Rb\Big)\Big\}\,
e^{\Delta_R\,(Y/2 - y)} \Big]\right\}.\nonumber
\eea
\begin{figure}
\begin{tabular}{c c}
\epsfig{file=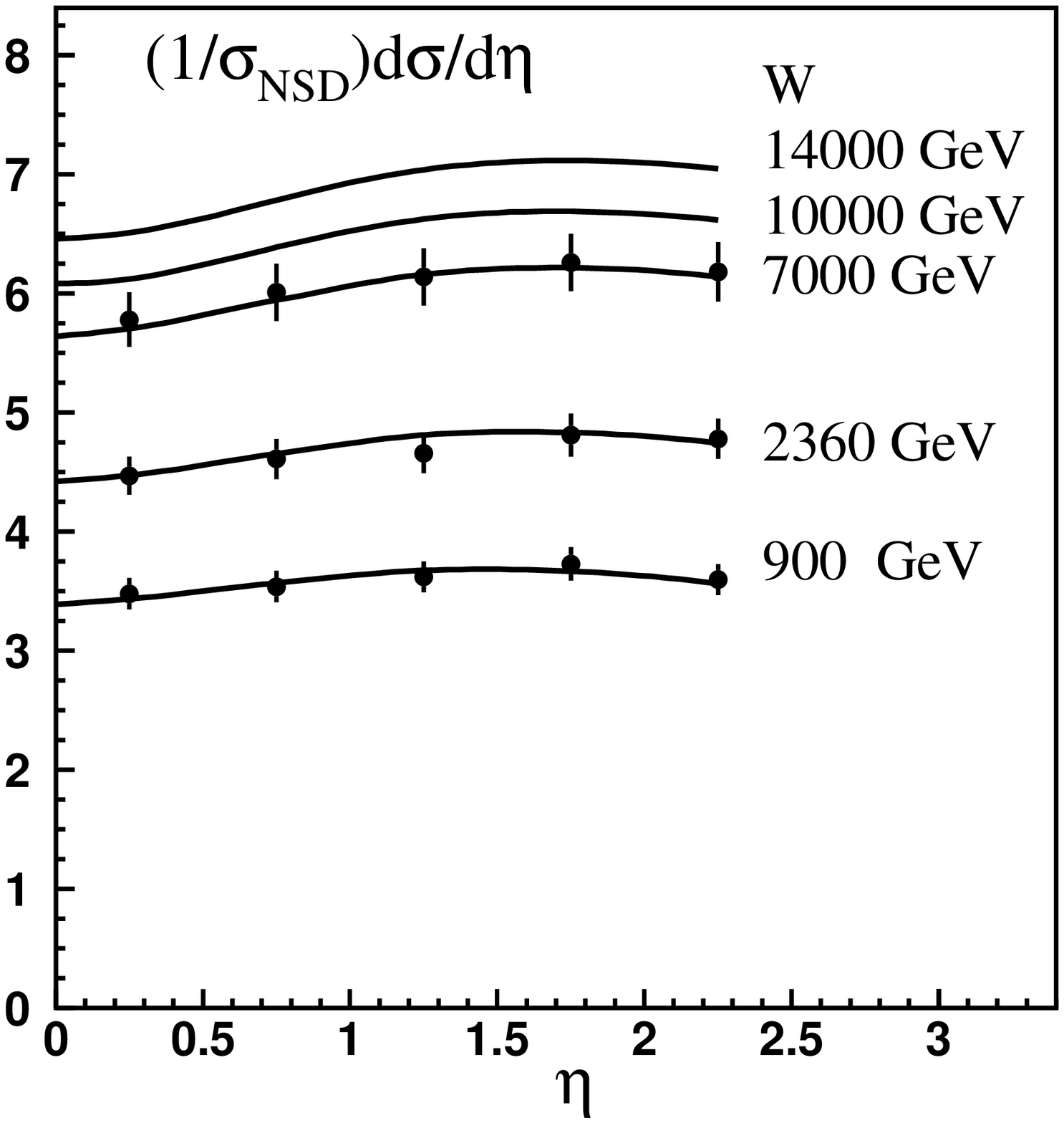,width=80mm} & \epsfig{file=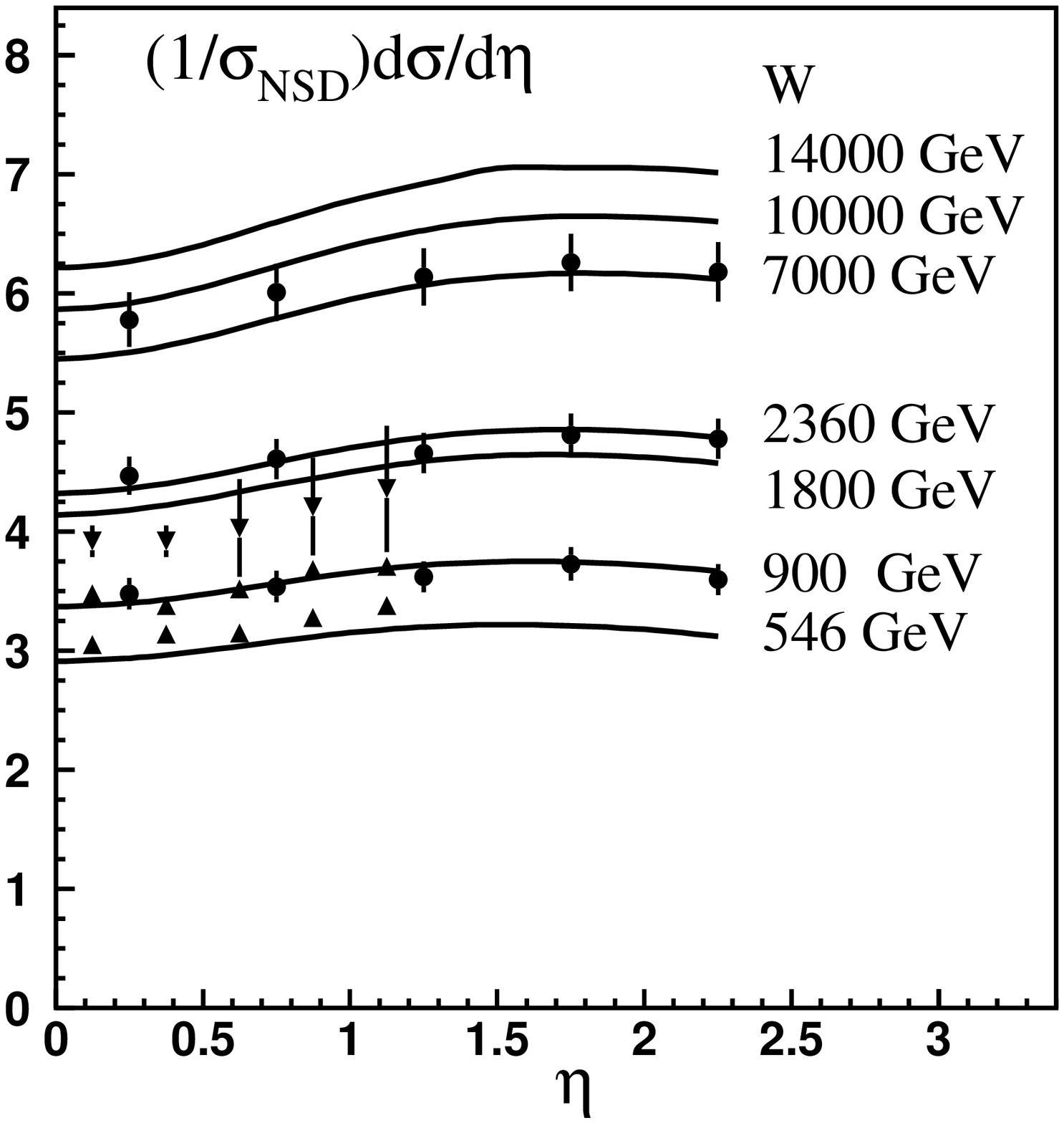,width=80mm}
\end{tabular}
\caption{The single inclusive density versus energy.  
The data were taken from Refs.\cite{ALICE,CMS,ATLAS} 
and from Ref.\cite{PDG}. 
The fit to the CMS data is plotted in \fig{incl}-a, 
while \fig{incl}-b presents the description of all inclusive spectra 
with $W \geq 546 GeV$.}
\label{incl}
\end{figure}

We extract the three new parameters: $a_{\pom \pom}, 
a_{\pom \reg}$ and $Q_0/Q$ from the experimental inclusive data. 
We made two separate fits: 
(a) fitting only the CMS data at different LHC
energies (see \fig{incl}-a); and (b) fitting all inclusive 
data for $W \geq 546 \,GeV$ (see \fig{incl}-b).
We choose only data in the central region of rapidity, 
as we have not included energy conservation, and therefore
our model is inadequate
to describe the data behavior in the fragmentation region. 
\begin{table}[ht]
\begin{minipage}{8cm}{
\begin{tabular}{|c|c|c|c|} \hline
Data & $a_{\pom \pom} $ &   $a_{\pom \reg}$  & $Q_0/Q$ \\
\hline \hline 
CMS  & 0.39  & 0.186 & 0.427\\ \hline 
All & 0.413 & 0.194 & 0.356\\ \hline \hline
\end{tabular}
}
\end{minipage}
\begin{minipage}{8cm}{
\caption{\label{table}
{Values of parameters for the fit of inclusive spectra}.}}
\end{minipage}
\end{table}
The values of fitted parameters are presented in the table. 
As stated, all 
other parameters were taken from Ref.\cite{GLMLAST}.

\fig{incl} shows that the soft model based on the Pomeron approach is able to
describe the behavior and the value of the inclusive production 
observed experimentally.
Our predictions are shown in the same figure. 
We note that the final version of our approach which includes
the contributions of enhanced, semi-enhanced and net diagrams 
(see Ref.\cite{GLMLAST}) provides a much better description of
the data than we obtained
in our previous attempt\cite{GLMINC}, 
where only enhanced diagrams were summed.

We believe that our description of the inclusive production 
presented here will be efficacious
in calculations of other observables at high
energies, such as correlations and multiplicity dependences.

Unfortunately, up to now, we are the only group that has attempted 
to describe 
inclusive production in the framework of a soft model. 
We hope that this effort 
will provide a background for other
microscopic approaches based on high density QCD.

\newpage
\centerline{\bf Acknowledgement}

We thank all participants of the conference “Diffraction 2010
” for fruitful discussion on the subject
The research of one of us (E.L.) was supported in part by the Fondecyt 
(Chile) grant \# 1100648.

\end{document}